\documentclass[12pt]{article}
\usepackage{amsmath, amsthm}
\usepackage{setspace}
\usepackage{amsfonts}
\numberwithin{equation}{section}
\usepackage[bottom]{footmisc}
\usepackage{fullpage}

\newcommand{\del}{\partial}

\title{
\hfill{\small WIS/9/11-NOV-DPPA \\}
\textmd{Effective String Theory and \\Nonlinear Lorentz \nolinebreak Invariance}
\vspace{5 mm}}
\author{Ofer Aharony$^1$\footnote{Ofer.Aharony@weizmann.ac.il}\,  and Matthew Dodelson$^{1,2}$\footnote{matthew\_dodelson@brown.edu}\vspace{5 mm}\\ \\
\emph{$^{1}$ Department of Particle Physics and Astrophysics}\\
\emph{Weizmann Institute of Science, Rehovot 76100, Israel}\vspace{3.5 mm}\\\emph{ $^2$ Department of Physics, Brown University}\\
\emph{Providence, RI 02912, USA}}
\date{
}
\begin{document}
\maketitle
\abstract{We study the low-energy effective action governing the transverse fluctuations of a long string, such as a confining flux tube in QCD. We work in the static gauge where this action contains only the
transverse excitations of the string. The static gauge action is strongly constrained by the requirement that the Lorentz symmetry, that is spontaneously broken by the long string vacuum, is nonlinearly realized on the Nambu-Goldstone bosons. One solution to the constraints (at the classical level) is the Nambu-Goto action, and the general solution contains higher derivative corrections to this. We show that in $2+1$ dimensions, the first allowed correction to the Nambu-Goto action is proportional to the squared curvature of the induced metric on the worldsheet. In higher dimensions, there is a more complicated allowed correction that appears at lower order than the curvature squared. We argue that this leading correction is similar to, but not identical to, the one-loop determinant $\sqrt{-h}R \Box^{-1}R$ computed by Polyakov for the bosonic fundamental string.}\thispagestyle{empty}
\pagebreak

\tableofcontents

\setcounter{page}{2}

\section{Introduction}

One-dimensional solitonic objects play an important role in many field theories, both strongly and weakly coupled -- examples include the Abrikosov-Nielsen-Olesen vortex \cite{Abrikosov,nielson} in the $3+1$ dimensional Abelian Higgs model, and confining color flux tubes connecting quark-antiquark pairs in quantum chromodynamics. In any such theory, the string breaks $D-2$ translations in a $D$-dimensional spacetime, so one universally expects $D-2$ massless scalars on the worldsheet by Goldstone's theorem. Furthermore, in the generic case where there are no additional symmetries, there is nothing to prevent the other modes on the string worldsheet from becoming massive, so the Nambu-Goldstone bosons are the only massless excitations of the string. It is then possible, at least in principle, to integrate out the massive fields in order to obtain a low-energy effective action for these Nambu-Goldstone modes, that is valid up to the mass of the lightest massive excitation. This action, which is valid for strings which fluctuate at long wavelengths compared to the other scales in the theory, is called the ``long string effective action.''

The choice of coordinates on the worldsheet of the string is arbitrary, so the effective action for the worldsheet embedding coordinates $X^{\mu}(\sigma^a)$ ($\mu=0,\cdots,D-1$, $a=0,1$) should be invariant under diffeomorphisms of $\sigma^a$. It is natural to go to a physical gauge for these diffeomorphism symmetries in which only the physical (transverse) fluctuations of the string appear in the action. For a string stretched predominantly along the $X^1$ direction in space (for instance, it could wrap a circle in this direction, or stretch between two boundaries localized in this direction), a natural choice is the static gauge $\sigma^0=X^0$, $\sigma^1=X^1$, which completely fixes the worldsheet diffeomorphism symmetry, and leaves only the transverse fluctuations $X^i(\sigma^a)$ ($i=2,\cdots,D-1$) which are the massless Nambu-Goldstone bosons. Such a choice is natural when expanding around the solution of a static long string stretched in the $X^1$ direction. The effective action in the static gauge is a functional of the $X^i$ and their derivatives, and has a low-energy expansion in the number of derivatives (terms involving $X^i$ with no derivatives cannot appear since the $X^i$ are Nambu-Goldstone bosons).

The disadvantage of the static gauge is that the space-time Lorentz symmetry, which is spontaneously
broken by the long string solution that we are expanding around, and by our static gauge choice, is
not manifest. However, this symmetry should still be non-linearly realized, and this leads to
constraints on the long string effective action. One obvious solution to these constraints is the
Nambu-Goto action
\begin{equation}\label{ngaction}
S_{\text{NG}} = - T \int d^2\sigma \sqrt{-h}~,
\end{equation}
where $T$ is the string tension, $h_{ab} \equiv \del_a X^{\mu} \del_b X_{\mu}$ is the induced
metric on the string worldsheet, and $h\equiv \det (h_{ab})$. This action is diffeomorphism-invariant and Lorentz-invariant, and
thus writing it in the static gauge automatically gives an action with a non-linearly realized
Lorentz symmetry. However, in general there could be more solutions to the constraints, which would
correct the Nambu-Goto action. In particular, any diffeomorphism-invariant functional of the induced metric, such as $\int d^2\sigma \sqrt{-h} R^n$ where $R$ is the induced curvature, is also automatically Lorentz-invariant when written in the static gauge, but it is not clear if such functionals are the only possible Lorentz-invariant actions or not.

The long string effective action was first systematically
analyzed in \cite{luscher}, up to four-derivative order. In this paper constraints on the effective action
were derived by requiring consistency (``open-closed duality'') between different interpretations of its partition function, involving propagation of the string in different channels. It was shown in \cite{luscher} that for $D=3$ this uniquely determined the coefficients of terms in the effective action with up to four derivatives up to an overall constant, implying that the action to this order
was equal to the Nambu-Goto action. It was later realized \cite{meyer,aharonykarzbrun,aharonykom} (see also \cite{Cohn:1992nu}) that the computation of \cite{luscher} assumed (through the form of the space-time
propagators) Lorentz symmetry, so that the constraints on the action really came from this symmetry.
The procedure of \cite{luscher} was subsequently generalized to any $D$ and carried out to the next order in the derivative expansion in \cite{aharonykarzbrun}. It was found that at six-derivative order a
correction to the Nambu-Goto action could appear with an arbitrary coefficient $c_4$, but only when
$D>3$, and it was verified that this is consistent with computations of the effective string action in a number of confining gauge theories with known weakly curved holographic duals. It was then noted in \cite{aharonykom}
that the constraint of Lorentz invariance could also be imposed by directly requiring that the action
is invariant under a non-linear Lorentz transformation of the transverse fields $X^i$, and that this gives
equivalent constraints (at least up to six-derivative order) to the ones found in \cite{aharonykarzbrun}. This method was then
used in \cite{aharonyfield} to analyze the leading corrections to the effective action for open strings.
The fact that the leading corrections to the string effective action appear at six-derivative order (for $D>3$) or at eight-derivative order (for $D=3$) implies that the deviations of the energy levels of long strings from their Nambu-Goto values are very small. The leading deviations were explicitly computed in
\cite{aharonykling}, and they are consistent with the latest lattice results for the spectrum of long confining strings (see \cite{Athenodorou:2011rx} and the references in \cite{aharonykling}, and see \cite{teper} for a recent review).\footnote{However, there is some unexplained tension between these results and lattice computations in some three dimensional models, see \cite{Giudice,Caselle,Billo':2010ix} and references therein.}

The non-linearly realized Lorentz transformation relates terms in the effective action of the
schematic form $d^n X^m$ to other terms with the same value of $(n-m)$. We call the value of $(n-m)$
the ``scaling'' of a given term. The terms of scaling zero were shown in \cite{JSZ,aharonykom,aharonyfield,dbi,Casalbuoni:2011fq} to be equivalent
to their Nambu-Goto value, to all orders in the derivative expansion. The leading
correction to Nambu-Goto found in \cite{aharonykarzbrun}, of the form $d^6 X^4$, is related by Lorentz
transformations to
other terms of scaling two, of the form $d^{2n+2} X^{2n}$ with $n=3,4,\cdots$. Up to now Lorentz
invariance was tested
order by order in the derivative expansion, and it is not clear if a given term (like the $c_4$
term mentioned above) has an all-orders Lorentz-invariant completion or not. This is obviously an
important question, since if there is no such completion for a given term then this term is not
allowed, leading to additional constraints on the effective action. In this paper we analyze
the Lorentz-invariance constraints on the terms with the lowest scaling that are allowed, to all
orders in the derivative expansion. For $D=3$ we find that the leading allowed eight-derivative term (of scaling four) has a unique Lorentz-invariant completion. For $D>3$ we argue (though we do not
rigorously prove) that again the leading allowed correction (the $c_4$ term) has a unique
Lorentz-invariant completion. This implies that the leading allowed corrections which were
assumed in the previous literature are indeed consistent with Lorentz invariance, at least classically.

Other gauge choices for the ``long string effective action'' can also be made. In particular,
in \cite{polstrom} the effective action was analyzed in the orthogonal gauge, in which the
induced metric is proportional to the Minkowski metric (see also \cite{polchinskiqcd,drummond,afk}). In this gauge Lorentz invariance is
manifest, but diffeomorphism invariance leads to non-trivial constraints. It was argued (but
not rigorously derived) in \cite{polstrom} that these constraints determine the leading
correction to the Nambu-Goto action in this gauge to take the form
\begin{equation}
\frac{\beta}{4\pi} \int d^2\sigma \sqrt{-h} R \frac{1}{\Box} R~,
\end{equation}
where $R$ is the curvature scalar of the induced worldsheet metric $h$ and $\beta = (26-D)/12$.
This is the same as Polyakov's one-loop determinant for the fundamental bosonic string,
but now written using the induced metric instead of an intrinsic worldsheet metric as in \cite{polyakov}.
The form that we will find for the leading correction to the action in $D>3$ will turn out to be
quite similar to this (but without the constraint on the coefficient), and we will discuss this further
below.

We begin in Section 2, by describing our general strategy and reviewing the form of the nonlinear Lorentz transformation and the equations of motion in the static gauge. In Sections 3 and 4, we derive the first allowed corrections to the Nambu-Goto action in $2+1$ dimensions and higher dimensions, respectively. We conclude in Section 5 with a summary of our results and possible directions for future investigation. Two appendices contain some technical details.

\section{Symmetries and Equations of Motion}

Consider the low-energy effective field theory on a string embedded in $D$ spacetime dimensions. The dynamical fields in the static gauge are the transverse coordinates $X^i(\sigma^0,\sigma^1)$
($i=2,\cdots,D-1$).
The full Poincar\'{e}-invariant field theory that our string is a solution of has a $SO(D-1,1)\times \mathbb{R}^D$ global symmetry, that is spontaneously broken in the long string vacuum $X^i=0$ to a $SO(D-2)\times SO(1,1)\times \mathbb{R}^2$ subgroup, consisting of rotations and boosts that do not mix the transverse and longitudinal fields, and translations along the worldsheet. One might expect that each generator that is not in this subgroup would correspond to a unique Nambu-Goldstone boson, but in fact the $D-2$ Nambu-Goldstone bosons for the broken translations are enough to realize the full Poincar\'{e} group \cite{low}. Since the effective action is not manifestly invariant under the broken rotations, these symmetries must be realized nonlinearly on the $X$ fields.

In order to derive the explicit form of this transformation, let us follow \cite{aharonykom, aharonyfield} and consider a broken infinitesimal boost $\delta_{02}$ and rotation $\delta_{12}$ in the $X^a-X^2$ plane, which act on the embedding coordinates before the gauge-fixing as
\begin{align}
&\delta_{02}X^a=\epsilon X^2\delta_0^a~,\hspace{15 mm}\delta_{02}X^i=\epsilon X^0\delta^i_2~,\\
&\delta_{12}X^a=\epsilon X^2\delta_1^a~,\hspace{15 mm}\delta_{12}X^i=-\epsilon X^1\delta^i_2~.
\end{align}
In both cases, the transformation of $X^a$ implies that the transformed field configuration is no longer in the static gauge, so we must make a compensating diffeomorphism $\delta_{a2}\sigma ^b=\epsilon X^2\delta^b_a$ on the worldsheet coordinates in order to leave our choice of gauge intact. Defining $\delta_{+2}=(\delta_{02}+\delta_{12})/\sqrt{2}$, the full transformation of the transverse fields
under this specific transformation is then
\begin{align}\label{lorentztrans}
&\notag\delta_{+2}(\partial_{+}X^i)=-\epsilon\partial_{+}(X^2\partial_+X^i)~,\\
&\delta_{+2}(\partial_{-}X^i)=\epsilon[\delta^i_2-\partial_{-}(X^2\partial_+X^i)]~,
\end{align}
where $\sigma^{\pm}=(\sigma^0\pm \sigma^1)/\sqrt{2}$ are light-cone coordinates.

In the following sections, we will find it useful to organize terms in the effective action by their scaling, which we define as the excess of derivatives over $X$ fields; for example, $(\partial^2_+X)^2(\partial_-X)^4$ has scaling two. The utility of this definition is that terms with different scaling do not mix under (\ref{lorentztrans}), so that one can individually analyze the part of the action containing terms with a fixed scaling. In fact, we will see that in some cases the term with a given scaling that is of lowest order in the derivative expansion determines (using Lorentz symmetry) all higher order terms with the same scaling.

By requiring the variation of the action under (\ref{lorentztrans}) to vanish, it was shown in \cite{aharonykom} that the scaling zero action is constrained to take the Nambu-Goto form (\ref{ngaction}) (as previously proven in \cite{JSZ}). In this work, we will therefore consider deviations from the Nambu-Goto action of the form $S=S_{\text{NG}}+\delta S$, where $\delta S$ has scaling greater than zero. Notice that $\delta S$ is small compared to $S_{\text{NG}}$, since we are working in a derivative expansion. Naively, we should now write down the most general possible $\delta S$, and require
that (like $S_{\text{NG}}$) $\delta S$ is also invariant under the transformation (\ref{lorentztrans}). However, if we consider the leading correction to the Nambu-Goto action, we can weaken this requirement in two ways.

First, we will allow variations of $\delta S$ under (\ref{lorentztrans}) that are proportional to the Nambu-Goto equations of motion. Such variations are still generated by currents that are conserved up to the Nambu-Goto equations of motion, and these are a good approximation to the full equations of motion of the theory,
so up to leading order in $\delta S$ this is enough for our purposes \footnote{More precisely, this is
enough to ensure that the Lorentz charges are still conserved at leading order in $\delta S$, but
their algebra could be modified \cite{aharonykom}.}.

Second, we can ignore terms in $\delta S$ that are proportional to the equations of motion, because these can be eliminated via field redefinitions. Note that a field redefinition would affect the form of the transformation (2.3) to first order in $\delta S$, but the only change in $\delta_{+2}S$ will be at $\mathcal{O}(\delta S^2)$, since the leading variation $\delta_{+2}S_{\text{NG}}$ is proportional (like any variation of the action) to the Nambu-Goto equations of motion. Again, at leading order in $\delta S$ we can drop terms proportional to the Nambu-Goto equations of motion rather than the full equations of motion of our action \footnote{From now on, whenever we refer to the equations of motion, we mean the Nambu-Goto equations of motion.}.

Let us now derive the form of the Nambu-Goto equations of motion in the static gauge. To do this, we first return to the Nambu-Goto theory in covariant form \eqref{ngaction}, where the equations of motion can be written as
\begin{align}\label{ngeom}
\partial_a(\sqrt{-h}h^{ab}\partial_bX^\mu)=0~,
\end{align}
where $h$ is the determinant of the induced metric $h_{ab} \equiv \del_a X^{\mu} \del_b X_{\mu}$, and $h^{ab}$ is its inverse.
Going to the static gauge, (\ref{ngeom}) becomes
 \begin{align}
&\partial_a(\sqrt{-h}h^{ab}\partial_bX^i)=0~, \label{eomone}\\
&\partial_a(\sqrt{-h}h^{ab})=0~. \label{eomtwo}
 \end{align}
One can check that (\ref{eomone}) implies (\ref{eomtwo}); this is required for the consistency of the fixing of the static gauge, since if the two equations were independent then the system would be overdetermined. Combining (\ref{eomone}) and (\ref{eomtwo}) gives $h^{ab}\partial_a\partial_bX^i=0$, or explicitly
\begin{align}\label{eom}
\partial_+\partial_-X^i=\frac{\partial_+^2X^i(\partial_-X)^2+\partial_-^2 X^i (\partial_+X)^2}{2(\partial_+X\cdot \partial_-X-1)}~.
\end{align}
The dot product here and below means a sum over the transverse index $i$, for instance $(\del_- X)^2 \equiv \del_- X^i \del_- X^i$.

\section{2+1 Dimensions}\label{threed}

The case of a string moving in $2+1$ dimensions (which is equivalent to a domain wall) is somewhat
simpler than the higher dimensional case, because in this case there is a single transverse coordinate $X$. This gives relations between various terms that differ in higher dimensions.
It is straightforward to check that no terms with scaling between one and three are allowed (all scaling two terms are total derivatives up to the equations of motion), so the first possible corrections to the Nambu-Goto action in this case arise at scaling four \cite{aharonykarzbrun}. The leading possible correction to Nambu-Goto in the derivative expansion takes the form $(\del_+^2 X)^2 (\del_-^2 X)^2$.

Up to integration by parts and up to the equations of motion, the most general $SO(1,1)$-invariant Lagrangian at scaling four that contains $2n$ $X$ fields takes the form
\begin{align}\label{lndthree}
\mathcal{L}_{4,n}&\notag= \left[a_n(\partial_+^3X)^2(\partial_+X)^{n-4}(\partial_-X)^{n+2}+b_n(\partial_+^2X)^4(\partial_+X)^{n-6}(\partial_-X)^{n+2}+(+\leftrightarrow -)\right]\\
&\hspace{4 mm}+c_{n}(\partial_+^2X\partial_-^2X)^2(\partial_+X\partial_-X)^{n-2}~.
\end{align}
We assume a worldsheet parity symmetry under $\sigma^+ \leftrightarrow \sigma^-$.
Each term in (\ref{lndthree}) is accompanied by negative powers of the tension. If our long string has a typical scale $\ell$ characterizing its length, we can rescale the worldsheet and space-time coordinates by this length, and then the derivative expansion in the worldsheet is an expansion in inverse powers of the
dimensionless parameter $T\ell^2$. In particular the energy levels of a string of length $\ell$ have an expansion of this form.

In order for the Lagrangian to be invariant under $\delta_{+2}$, the quantity $\sum_n \delta_{+2}\mathcal{L}_{4,n}$ is required to vanish. After using the equations of motion, a basis for the terms appearing in $\sum_n\delta_{+2}\mathcal{L}_{4,n}$ that are linearly independent up to integration by parts is given by
\begin{align}
(\partial_+^3X)^2~,\hspace{10 mm}(\partial_+^2 X)^4~,\hspace{10 mm}(\partial_+^2X)^2(\partial_-^2X)^2~,
\end{align}
times appropriate powers of $(\del_+X)$ and $(\del_-X)$, and their counterparts with $+\leftrightarrow -$. Varying (\ref{lndthree}), one finds that only the variation of the $a_n$ term yields terms involving $(\partial_+^3X)^2$, implying that $a_n=0$ for all $n$.
The variation of the remaining terms in (\ref{lndthree}) can be expanded in terms of $(\partial_{\pm}^2 X)^4$ and $(\partial_+^2X)^2(\partial_-^2X)^2$, giving two relations between $b_{n},b_{n-1},c_n, $ and $c_{n-1}$ for every value of $n$. This suffices to uniquely determine all coefficients in the action up to one overall constant. Solving the recursion relation is best done using a trick, and we leave the details to an appendix.

Up to an overall normalization, the unique solution to the Lorentz-invariance condition at scaling four may be written in the diffeomorphism-invariant form
\begin{align}\label{rsquared}
\delta {\cal L} \propto \sqrt{-h}R^2\propto \frac{\left[\partial_+^2X\partial_-^2X-(\partial_+\partial_-X)^2\right]^2}{\left(1-2\partial_+X\partial_-X\right)^{7/2}}~,
\end{align}
where $R$ is the scalar curvature constructed from the induced worldsheet metric $h$. 
As expected for a diffeomorphism-invariant term, one can check that (\ref{rsquared}) is invariant under the nonlinear Lorentz transformation even off-shell (without using the equations of motion).

 The reader may be wondering why our analysis did not identify the Euler characteristic $\int d^2\sigma\, \sqrt{-h}R$, which has scaling two, as being invariant under the Lorentz transformation\footnote{We thank M. Field for clarifying discussions on this point.}. This term is a topological invariant and does not affect the equations of motion, but it weights amplitudes by a factor related to the genus of the worldsheet, analogously to the dilaton-curvature coupling in string theory \cite{polchinski}. In our long string expansion we do not
 allow any topologically non-trivial worldsheets so this term should be trivial. Indeed, one finds that $\sqrt{-h}R$ is a total derivative as long as its Taylor series in the $\partial X$'s converges, which is the case for the long string expansion in the static gauge.

\section{Higher Dimensions}

In more than three dimensions, there are several $X$'s, and more general terms, including terms of scaling two, may also be written \cite{aharonykarzbrun}. The general $SO(D-2)\times SO(1,1)$-invariant action with $n$ pairs of $X$ fields at scaling two is
\begin{align}\label{genterm}
\mathcal{L}_{2,n}\notag&=\left[(\partial_+^2 X\cdot \partial_+X)^2((\partial_-X)^2)^3\mathcal{L}_{0,n-5}^{\text{a}}+(\partial_+^2 X\cdot \partial_+ X)(\partial^2_+X\cdot \partial_-X)((\partial_-X)^{2})^2\mathcal{L}^{\text{b}}_{0,n-4}\right.\\
&\left.\notag\hspace{4.5 mm}+(\partial^2_+X)^2((\partial_- X)^2)^2\mathcal{L}_{0,n-3}^{\text{c}}+(\partial_+^2 X\cdot \partial_-X)^2(\partial_-X)^2\mathcal{L}^{\text{d}}_{0,n-3}+(+\leftrightarrow -)\right]\\
&\hspace{4.5 mm}+(\partial_+^2 X\cdot \partial_+ X)(\partial^2_-X\cdot \partial_-X)\mathcal{L}^{\text{e}}_{0,n-2}+(\partial_+^2 X\cdot \partial^2_-X)\mathcal{L}^{\text{f}}_{0,n-1}~,
\end{align}
where the scaling zero Lagrangian with $2n$ $X$ fields is defined by
\begin{align}
\mathcal{L}_{0,n}=\sum_{m}a_{n,m}((\partial_+X)^{2})^m((\partial_- X)^{2})^m(\partial_+ X\cdot \partial_- X)^{n-2m}~.
\end{align}
The different superscripts in (\ref{genterm}) indicate that each $\mathcal{L}_{0,n}$ involves different coefficients \nolinebreak $a_{n,m}$.

Let us now sketch the method for varying (\ref{genterm}). After applying $\delta_{+2}$ using (\ref{lorentztrans}), we integrate by parts to put every term into the form $X^2(\cdots)$. Then, we use the equations of motion (\ref{eom}) to eliminate terms proportional to $\partial_+\partial_-X$ or its derivatives. Some of the remaining terms are not linearly independent: for example, up to total derivatives, we have
\begin{align}
X^2(\partial_+^3X\cdot \partial_-^2X)&=-X^2\left\{\partial_+\left(\partial_-X\cdot \partial_+^2\partial_-X\right)+\partial_+^2X\cdot \partial_+\partial_-^2X\right.\\
&\hspace{-10 mm}\left.\notag+\frac{1}{2}\partial_+^2\left[\frac{(\partial_+^2X\cdot \partial_-X)(\partial_-X)^2}{1-\partial_+X\cdot \partial_-X}\right]+\frac{1}{2}\partial_-^2\left[\frac{(\partial_+^2X\cdot \partial_-X)(\partial_+X)^2}{1-\partial_+X\cdot \partial_-X}\right]\right\}~.
\end{align}
After getting rid of such  terms, one must iterate the above procedure in order to find the full variation. Sparing the reader from the remainder of the details, the solution up to 16-derivative order is unique up to an overall constant, and may be written in the form
\begin{align}\label{result}
\delta \mathcal{L} = 2c_4 & \sqrt{-h}R\left[\log(\sqrt{-h})-\frac{(\partial_+X)^2(\partial_-X)^2}{4(1-\partial_+X\cdot \partial_-X)^2}-\frac{5((\partial_+X)^2)^2((\partial_-X)^2)^2}{32(1-\partial_+X\cdot \partial_-X)^4}\right.\\
&\notag\left.\hspace{13 mm}-\frac{11((\partial_+X)^2)^3((\partial_-X)^2)^3}{96(1-\partial_+X\cdot \partial_-X)^6}-\frac{93((\partial_+X)^2)^4((\partial_-X)^2)^4}{1024(1-\partial_+X\cdot \partial_-X)^8}\right]+\mathcal{O}((T\ell^2)^{-10})~,
\end{align}
where we normalized $c_4$ so that it agrees with the literature \cite{aharonykling}. Note that this expression is a total derivative in three dimensions, which is consistent with the fact that there are no allowed scaling two terms in $D=3$.

We were not able to solve the Lorentz variation requirements explicitly to all orders in the derivative
expansion. If we could write our expression in a diffeomorphism invariant form, say as some functional of
the induced curvature $R$, then it would be clear how to do this, but we could not write (\ref{result}) directly in such a form. However, we can
express (\ref{result}) in a more transparent form. To do this, let us define an operator $\tilde{\Box}^{-1}$ by the relation
\begin{align}\label{boxdef}
\frac{1}{\tilde{\Box}}f=\frac{1}{{\Box}}(f+\text{equations of motion})~.
\end{align}
That is, to compute $\tilde{\Box}^{-1}f$, one adds a function to $f$ that is proportional to the equations of motion, such that the combination may be written in the form $\Box g$ for some function $g$, and takes ${\tilde{\Box}}^{-1} f = g$.
For general $f$ this definition is plagued with ambiguities -- however, we show in an appendix that $\tilde{\Box}^{-1}R$ is uniquely defined. The existence of $\tilde{\Box}^{-1}R$ is a more difficult question; we could not prove this, but it is straightforward to check that it holds order by order in the derivative expansion. Next, note that, up to the equations of motion, the worldsheet Laplacian is equal to
\begin{equation}
\Box= h^{ab}\partial_a\partial_b
=\frac{1}{h}\left[(\partial_+X)^2\partial_-^2+(\partial_-X)^2 \partial_+^2+2(1-\partial_+X\cdot \partial_-X)\partial_+\partial_-\right]~,
\end{equation}
where the first equality follows from (\ref{eomtwo}). Using this form of the Laplacian, one can then check that (\ref{result}) can be rewritten as
\begin{align}\label{simplified}
\delta \mathcal{L}= -2c_4 \sqrt{-h}R\frac{1}{\tilde{\Box}}R+\mathcal{O}((T\ell^2)^{-10})~.
\end{align}

Motivated by (\ref{simplified}), we conjecture that $\sqrt{-h}R\tilde{\Box}^{-1}R$ is invariant under the Lorentz transformation to all orders in the derivative expansion, and is therefore the only allowed correction to the Nambu-Goto action at scaling two. The major obstacle to proving this directly is that we could not systematically compute $\tilde{\Box}^{-1}R$. We expect that $\tilde{\Box}^{-1}R$  can be expressed as an infinite series whose radius of convergence is on the order of $T\ell^2$, where the effective field theory breaks down, but we have not yet been able to identify this series.

One may expect that the leading correction to the action would take the form $\sqrt{-h} R {\Box}^{-1} R$, as found in the orthogonal gauge in \cite{polstrom}. This term is manifestly Lorentz-invariant, but it does not seem to be local in the static gauge, in the sense of having a good derivative expansion.
Note that $\sqrt{-h}R\tilde{\Box}^{-1}R$ is not equivalent to $\sqrt{-h}R{\Box}^{-1}R$ under a field redefinition, since $\Box^{-1}$ acting on the equations of motion in (\ref{boxdef}) is not proportional to the equations of motion. However, if we use the same definition (\ref{boxdef}) in the orthogonal gauge, the two terms would be equivalent there. Perhaps $\tilde{\Box}^{-1}$ should be thought of as some regularized form of the inverse Laplacian, but it is not clear what regularization scheme is being used, since the subtractions $\Box^{-1}(\text{equations of motion})$ are non-local.

\section{Conclusions}

We have used a nonlinear realization of Lorentz symmetry to constrain deviations of the static gauge effective action on a long string from the Nambu-Goto action. Combined with the known results at scaling zero, our analysis implies that the action must take the form
\begin{align}
S=\begin{cases}
&-T\int d^2\sigma\sqrt{-h}\left(1+aR^2+\text{higher scaling}\right)\qquad\text{ for $D=3$}\\
&-T\int d^2\sigma\sqrt{-h}\left(1+\frac{2c_4}{T} R\tilde{\Box}^{-1}R+\text{higher scaling}\right)\text{ for $D>3$}~,
\end{cases}
\end{align}
where $a$ and $c_4$ are arbitrary constants, and $\tilde{\Box}^{-1}$ was defined in (\ref{boxdef}). The consistency of the second term in $D>3$ has not been proven, but we have tested it to high orders in derivative expansion. This result confirms that the leading possible corrections in the derivative expansion that were discussed in \cite{aharonykarzbrun,aharonykom,aharonykling} are allowed by Lorentz symmetry, at least at the classical level.

One of the most interesting open questions is the relation between our results and those of \cite{polstrom}. Since we cannot write our correction in a manifestly diffeomorphism-invariant fashion, we cannot directly compare it to other gauges, such as the gauge used in \cite{polstrom}. However, one can check \cite{afk} that computations of gauge-invariant quantities like energy levels using the effective actions we found here agree with the results using the formalism of \cite{polstrom}. It was claimed in \cite{polstrom} that quantum considerations fix the coefficient of the leading correction uniquely. In our classical analysis this coefficient $c_4$ is arbitrary, and it would be interesting to understand if quantum corrections constrain it somehow. These issues will be discussed further in \cite{aharonykom}.

There are various possible generalizations of our computations.
In this paper we only discussed closed strings, with no boundary terms.
 Boundary terms can also be analyzed using methods similar to those presented here \cite{aharonyfield}, and it would be interesting to explicitly solve the all-order constraints on the boundary terms for low values of the scaling.
Similarly, one can generalize our considerations to include gauge fields, as they appear in D-brane actions; the scaling zero action was shown in \cite{dbi,Casalbuoni:2011fq} to agree with the Dirac-Born-Infeld action, and it would be interesting to analyze the leading correction to this. In our analysis we assumed worldsheet and space-time parity, and we did not include any terms involving space-time Levi-Civita tensors in our analysis. It would be interesting to understand if such terms, suggested for instance in \cite{mazur},  can also arise in the long string effective action.

One may also wonder about the consequences of adding more massless fields to the worldsheet effective field theory. For example, a confining string in a supersymmetric gauge theory generically breaks all of the supersymmetries, giving massless Goldstinos on the worldsheet. The effective action is therefore constrained by nonlinearly realized supersymmetry, and it is tempting to predict that the resulting constraints would imply that the first correction to the Ramond-Neveu-Schwarz action is related to the supersymmetric Liouville theory determinants calculated in \cite{superstring} (in the same sense that
the correction we found is related to the bosonic Liouville action found by Polyakov).


\section*{Acknowledgements} MD would like to thank the Weizmann high energy theory group for their hospitality during the course of this work, M. Field for several enlightening discussions, and S. Dodelson for helpful comments on the manuscript. OA would like to thank M. Field, N. Klinghoffer, Z. Komargodski and A. Schwimmer for collaboration on related topics, and for many useful discussions and comments on this manuscript. MD's research was supported by the Kupcinet-Getz Summer School at the Weizmann Institute of Science. The work of OA was supported in part by the Israel--U.S.~Binational Science Foundation, by a research center supported by the Israel Science Foundation (grant number 1468/06), by the German-Israeli Foundation (GIF) for Scientific Research and Development, and by the Minerva foundation with funding from the Federal German Ministry for Education and Research.

\appendix

\section{The Lorentz Variation in $D=3$}

Instead of directly extracting the constraints on the $D=3$ Lagrangian (\ref{lndthree}) from its variation, we will add some terms to the action that are proportional to the equations of motion in order to make the variation of the action simpler. A combination of a term of the form
\begin{align}\label{polynomial}
[\alpha_n(\partial_+\partial_-X)^2+\beta_n\partial_+^2X\partial_-^2X](\partial_+\partial_-X)^2
(\partial_+X\partial_-X)^{n-2}
\end{align}
(for arbitrary $\alpha_n$ and $\beta_n$) with the $b_n$ and $c_n$ terms in (\ref{lndthree}) is proportional to the equations of motion, so it makes no difference to add it to the action, as long as we shift $b_n$
and $c_n$ accordingly (in this appendix we will use the shifted $b_n$ and $c_n$ everywhere).
Setting $a_n=0$
as found in section \ref{threed}, the variation of the action (\ref{lndthree})+(\ref{polynomial}) then becomes (after integrations by parts)
\begin{align}\label{lorentzvar}
&\delta_{+2}\mathcal{L}_{4,n}=-\epsilon(\partial_+X)^{n-2}(\partial_-X)^{n-3} \left\{\left[2(2c_n+\beta_n) (\partial_+^2X)^2\partial_-^2X\partial_+\partial_-X\right.\right.\\
&\left.\notag\hspace{56 mm}+2(2\alpha_n+\beta_n)\partial_+^2X(\partial_+\partial_-X)^3\right](\partial_-X)^{2}\\
&\left.\notag\left.+\left[(2n+3)\partial_+X\partial_-X+2-n\right]\left[\alpha_n(\partial_+\partial_-X)^4+\beta_n(\partial_+\partial_-X)^2\partial_+^2X\partial_-^2X+c_n(\partial_+^2X\partial_-^2X)^2\right]\right.\right\}\\
&\notag+\epsilon b_n\left\{(\partial_+^2X)^4(\partial_+X)^{n-6}(\partial_-X)^{n+1}\left[n+2-(2n+7)\partial_+X\partial_-X\right]\right.\\
&\notag\hspace{10 mm}\left.+(\partial_-^2X)^3(\partial_+X)^{n+2}(\partial_-X)^{n-7}\left[(n-6)\partial_-^2X+(1-2n)\partial_-^2X\partial_+X\partial_-X-8\partial_+\partial_-X(\partial_-X)^2\right]\right\}~.
\end{align}

If we do not use the equations of motion on (\ref{lorentzvar}), then the sum $\sum_n \delta_{+2}\mathcal{L}_{4,n}$ vanishes if and only if $b_n=0$ and $\alpha_n=-\beta_n/2=c_n$,
and in addition we get a recursion relation for the $c_n$ :
\begin{align}
c_{n}=\frac{2n+1}{n-2}c_{n-1}
\end{align}
for $n\ge 3$. The solution to this recursion relation is
\begin{align}\label{solution}
\alpha_n=-\frac{\beta_n}{2}=c_{n}\propto \frac{(2n+1)!!}{(n-2)!}
\end{align}
 for $n\ge 2$, which reproduces the Taylor expansion of (\ref{rsquared}), as claimed. In fact, one can check that (\ref{solution}) is the unique solution to the constraints even if we allow (\ref{lorentzvar}) to be proportional to the equations of motion.

\section{Uniqueness of $\tilde{\Box}^{-1}R$}

In order to show that (\ref{simplified}) is well-defined, we must check that there is a unique solution to $\Box g=R+(\text{equations of motion})$ up to shifts of $g$ by terms proportional to the equations of motion, as long as $g$ is assumed to have a good expansion in derivatives.

To prove this, first note that once the equations of motion are used, $g$ must be a sum of terms of the schematic form $(\partial X)^n$. Indeed, $g$ has scaling zero, and each $X$ field in $g$ must be differentiated; if not, then $R$ would necessarily contain terms where $X$ is not differentiated, which is not the case.
The $SO(D-2)\times SO(1,1)$ symmetry then
implies that $g$ is a sum of terms of the form
\begin{align}
g_{n,m}=(\partial_+X\cdot \partial_-X)^{n}((\partial_+X)^{2})^m((\partial_-X)^{2})^m~.
\end{align}
It is then straightforward to check that the functions $\Box g_{n,m}$ are linearly independent after use of the equations of motion, so there is enough information in the equation $\Box g=R+(\text{equations of motion})$ to uniquely determine the coefficient of each $g_{n,m}$. This completes the argument.


{\footnotesize

\begin{thebibliography}{99}
\bibitem{Abrikosov}
  A.~A.~Abrikosov,
  ``On the Magnetic properties of superconductors of the second group,''
  Sov.\ Phys.\ JETP {\bf 5 } (1957)  1174.
  \bibitem{nielson}
H. B. Nielsen and P. Olesen, ``Vortex-line models for dual strings," Nucl. Phys. B \textbf{61} (1973) 45.
 \bibitem{luscher}
M. L\"{u}scher and P. Weisz, ``String excitation energies in SU(N) gauge theories beyond the free-string approximation," JHEP \textbf{07} (2004) 014, {\tt hep-th/0406205}.
\bibitem{meyer}
  H.~B.~Meyer,
  ``Poincare invariance in effective string theories,''
  JHEP {\bf 0605} (2006) 066
  [arXiv:hep-th/0602281].
\bibitem{aharonykarzbrun}
O. Aharony and E. Karzbrun, ``On the effective action of confining strings," JHEP \textbf{0906} (2009) 012, arXiv:0903.1927v4 [hep-th].
\bibitem{aharonykom}
O. Aharony, Z. Komargodski, and A. Schwimmer, work in progress, presented by O.~Aharony at the Strings 2009 conference, June 2009,
 {\tt http://strings2009.roma2.infn.it/talks/Aharony\_Strings09.ppt}, and
 at the ECT* workshop on ``Confining flux tubes and strings'', July 2010,
 {\tt http://www.ect.it/Meetings/ConfsWksAndCollMeetings/ConfWksDocument/
 2010/talks/Workshop\_05\_07\_2010/Aharony.ppt}.
\bibitem{Cohn:1992nu}
  J.~D.~Cohn, V.~Periwal,
  ``Lorentz invariance of effective strings,''
  Nucl.\ Phys.\  {\bf B395 } (1993)  119-128.
  [hep-th/9205026].
\bibitem{aharonyfield}
O. Aharony and M. Field, 	``On the effective theory of long open strings," JHEP \textbf{1101} (2011) 065, 	 arXiv:1008.2636v2 [hep-th].
\bibitem{aharonykling}
O. Aharony and N. Klinghoffer, ``Corrections to Nambu-Goto energy levels from the effective string action," JHEP \textbf{1012} (2010) 058, arXiv:1008.2648v2 [hep-th].
\bibitem{Athenodorou:2011rx}
  A.~Athenodorou, B.~Bringoltz, M.~Teper,
  ``Closed flux tubes and their string description in D=2+1 SU(N) gauge theories,''
  JHEP {\bf 1105 } (2011)  042.
  [arXiv:1103.5854 [hep-lat]].
\bibitem{teper}
M. Teper, ``Large N and confining flux tubes as strings -- a view from the lattice," arXiv:0912.3339 [hep-lat].
\bibitem{Giudice}
  P.~Giudice, F.~Gliozzi and S.~Lottini,
  ``The confining string beyond the free-string approximation in the gauge dual
  of percolation,''
  JHEP {\bf 0903} (2009) 104
  [arXiv:0901.0748 [hep-lat]].
\bibitem{Caselle}
  M.~Caselle and M.~Zago,
  ``A new approach to the study of effective string corrections in LGTs,''
  Eur.\ Phys.\ J.\  C {\bf 71} (2011) 1658
  [arXiv:1012.1254 [hep-lat]].
\bibitem{Billo':2010ix}
  M.~Billo', M.~Caselle, V.~Verduci, M.~Zago,
  ``New results on the effective string corrections to the inter-quark potential,''
  PoS {\bf LATTICE2010 } (2010)  273.
  [arXiv:1012.3935 [hep-lat]].
\bibitem{JSZ}
  S.~Jaimungal, G.~W.~Semenoff and K.~Zarembo,
  ``Universality in effective strings,''
  JETP Lett.\  {\bf 69} (1999) 509
  [arXiv:hep-ph/9811238].

  \bibitem{dbi}
  F. Gliozzi, ``Dirac-Born-Infeld action from spontaneous breakdown of Lorentz symmetry in brane-world scenarios," Phys. Rev. D \textbf{84} (2011) 027702, arXiv:1103.5377v4 [hep-th].
\bibitem{Casalbuoni:2011fq}
  R.~Casalbuoni, J.~Gomis, K.~Kamimura,
  ``Space-time transformations of the Born-Infeld gauge field of a D-brane,''
  [arXiv:1104.4916 [hep-th]].
\bibitem{polstrom}

 J. Polchinski and A. Strominger, ``Effective string theory," Phys. Rev. Lett. \textbf{67} (1991) 1681.

 \bibitem{polchinskiqcd}
 J. Polchinski, ``Strings and QCD?" 	{\tt hep-th/9210045}.
  \bibitem{drummond}
  J. M. Drummond, ``Universal subleading spectrum of effective string theory," {\tt hep-th/0411017}.
  \bibitem{afk}
    O. Aharony, M. Field, N. Klinghoffer, ``The effective string spectrum in the orthogonal gauge," to appear.

  \bibitem{polyakov}
  A. M. Polyakov, ``Quantum geometry of bosonic strings," Phys. Lett. B \textbf{103} (1981) 207.
  \bibitem{low}
  I. Low and A. V. Manohar, ``Spontaneously broken spacetime symmetries and Goldstone's theorem,"  Phys. Rev. Lett. \textbf{88} (2002) 101602, {\tt hep-th/0110285}.

  \bibitem{polchinski}
 J. Polchinski, \emph{String Theory}, Vol. 1, Cambridge University Press (1998).

  \bibitem{mazur}
  P. O. Mazur and V.P. Nair, ``Strings in QCD and $\theta$-vacua," Nucl. Phys. B \textbf{284} (1987) 146.
      \bibitem{superstring}
    A. M. Polyakov, ``Quantum geometry of fermionic strings," Phys. Lett. B \textbf{103} (1981) 211.


  \end{thebibliography}}
\end{document}